\documentclass{jfm}

\usepackage{graphicx}
\usepackage{epstopdf, epsfig}
\usepackage[hidelinks]{hyperref}

\shorttitle{Self-similar geometries within the inertial subrange}
\shortauthor{M. Heisel, C. M. de Silva, G. G. Katul, and M. Chamecki}

\title{Self-similar geometries within the inertial subrange of scales in boundary layer turbulence}

\author{Michael Heisel\aff{1}
  \corresp{\email{heisel@ucla.edu}},
  Charitha M. de Silva\aff{2},
  Gabriel G. Katul\aff{3,4},
  \and Marcelo Chamecki\aff{1}}

\affiliation{\aff{1}Department of Atmospheric and Oceanic Sciences, University of California in Los Angeles, Los Angeles, CA 90095, USA
\aff{2}School of Mechanical and Manufacturing Engineering, University of New South Wales, Sydney 2052, Australia
\aff{3}Department of Civil and Environmental Engineering, Duke University, Durham, NC 27708, USA
\aff{4}Nicholas School of the Environment, Duke University, Durham, NC 27708, USA}

\begin{document}

\maketitle

\begin{abstract}

The inertial subrange of turbulent scales is commonly reflected by a power law signature in ensemble statistics such as the energy spectrum and structure functions -- both in theory and from observations. Despite promising findings on the topic of fractal geometries in turbulence, there is no accepted image for the physical flow features corresponding to this statistical signature in the inertial subrange. The present study uses boundary layer turbulence measurements to evaluate the self-similar geometric properties of velocity isosurfaces and investigate their influence on statistics for the velocity signal. The fractal dimension of streamwise velocity isosurfaces, indicating statistical self-similarity in the size of ``wrinkles'' along each isosurface, is shown to be constant only within the inertial subrange of scales. For the transition between the inertial subrange and production range, it is inferred that the largest wrinkles become increasingly confined by the overall size of large-scale coherent velocity regions such as uniform momentum zones. The self-similarity of isosurfaces yields power law trends in subsequent one-dimensional statistics. For instance, the theoretical 2/3 power law exponent for the structure function can be recovered by considering the collective behavior of numerous isosurface level sets. The results suggest that the physical presence of inertial subrange eddies is manifested in the self-similar wrinkles of isosurfaces.

\end{abstract}

\section{Introduction}
\label{section1}

The sheer complexity of turbulence leads to the description of its features in abstract terms such as ``eddies''. For instance, texts often resort to amorphous forms referred to as ``blobs'', ``parcels'', and ``soups'' to illustrate theoretical constructs \citep{Tennekes1972,Frisch1995,Davidson2015}. In the last few decades, research efforts in the area of coherent structures has given more definitive shape to persistent turbulent features \citep{Cantwell1981,Robinson1991,Jimenez2018}. The smallest features associated with the dissipative range of scales -- often identified as coherent regions of velocity gradient statistics -- resemble sheets and tubes of lower-magnitude vorticity, worm-like filaments of intense vorticity, and sheets of intense dissipation \citep[e.g.,][]{Batchelor1949,Kuo1972,She1990,Jimenez1993,Vincent1994,Moisy2004,Elsinga2010}. These shapes, akin to rods and slabs, are directly related to the principle invariants of the velocity gradient tensor acting to deform the local fluid.

The physical representation of larger-scale features is specific to the flow configuration. For boundary layer turbulence, identified geometries include packets of hairpin-shaped structures \citep{Head1981}, elongated streak-like structures of coherent velocity \citep{Hwang2015} that tend to meander \citep{Kevin2019,deSilva2020}, and weakly-rotating streamwise rolls \citep{delAlamo2006} coinciding with side-by-side low- and high-momentum streaks \citep{Dennis2011}. Within the span of turbulent scales, these features correspond to the production range exhibiting a $-1$ power law exponent in the energy spectrum \citep{Tchen1953,Perry1982,Perry1986,Nickels2005,Katul2012,Calaf2013} and (very-)large-scale motions whose length exceeds the boundary layer thickness \citep{Kim1999,Guala2006,Balakumar2007,Smits2011,Lee2014}.

The spatial organization of these small- and large-scale features is apparent in the framework of uniform momentum zones (UMZs) \citep{Meinhart1995}. In this framework, the outer region of high-Reynolds-number boundary layers is approximated as a population of relatively uniform streamwise velocity regions, i.e. UMZs \citep{Adrian2000,deSilva2016}. The generic definition of UMZs can encompass more specific coherent features discussed above. For instance, near-wall UMZs bear the signature of three-dimensional streaks in the buffer and logarithmic regions \citep{Hwang2018,Cheng2020,Bae2021} and UMZs farther from the wall resemble bulges in the outer region of the boundary layer \citep{Kovasznay1970,Falco1977}. The size and velocity scaling behavior of UMZs is directly related to ensemble statistics such as the logarithmic mean velocity profile \citep{Heisel2020}. Further, the most intense small-scale features such as vortex cores are intermittently distributed and often reside along thin high-shear layers that align with the interfaces between UMZs \citep{Eisma2015,deSilva2017,Heisel2018,Bautista2019,Gul2020,Heisel2021}.

Despite the progress characterizing the size and shape of coherent flow regions, there remains a knowledge gap pertaining to intermediate eddies. The large-scale velocity regions and small dissipative features described above do not account for the inertial subrange of turbulent scales that occupies the Fourier space between the integral and dissipative motions \citep{Kolmogorov1941,Obukhov1941,Kolmogorov1962}. This region is characterized by power law exponents --5/3 and 2/3 in the energy spectrum and second-order structure function, respectively \citep{Kolmogorov1941}. Perhaps the most successful endeavor to identify inertial subrange eddies in physical space comes from the study of fractal geometries in turbulence \citep{Mandelbrot1974,Frisch1978,Mandelbrot1982}. Fractal isosurfaces and interfaces are naturally consistent with the signature of self-similarity across the inertial scales and these geometries have been observed for a variety of turbulent flows \citep[][among others]{Sreenivasan1986,Meneveau1991,Sreenivasan1991,Constantin1991,Brandenburg1992,Moisy2004,Lozano2012,deSilva2013,Borrell2016}. Yet, many of the observations are limited to coarse approximations of fractal attributes due to the requirement for high-fidelity measurements in flows with sufficient scale separation (i.e. high Reynolds number) for an extensive self-similar range of scales to develop. In particular, the estimated fractal dimension can appear scale-dependent due to both experimental factors and Reynolds-number effects \citep{Iyer2020,Catrakis1996,Catrakis2000,Heisel2022}, which may explain conflicting findings on the existence of monofractals in turbulence \citep[e.g.,][]{Miller1991,Praskovsky1993,Villermaux1999}. In consideration of these challenges, it has not been conclusively shown that the fractal geometry of isosurfaces in turbulence coincides specifically with the inertial subrange, despite suggested associations \citep[e.g.,][]{Sreenivasan1986,Sreenivasan1989}.

To this end, the topic of fractal geometries in turbulence is revisited here using high-Reynolds-number boundary layer measurements. The analysis uses the framework of UMZs to evaluate fractal geometries in the context of recent advances on the topic of coherent flow structures. In particular, \citet{deSilva2017} observed that streamwise velocity isosurfaces representing the edges of UMZs exhibit self-similar fractal properties. The present work expands on this finding by assessing the geometric properties of UMZs with a focus on the inertial subrange of scales. The study seeks to elucidate how flow features in physical space reflect the statistical signature of inertial subrange eddies. The term ``self-similar'' is used here to describe geometries whose fine- and coarse-scaled features have the same statistical shape properties, which yields power-law behavior in scale-dependent statistics. This meaning is distinct from self-similarity in the overall size and aspect ratio of large-scale features across varying wall-normal distance, which is also a characteristic of boundary layer turbulence \citep[e.g.,][]{delAlamo2006,Lozano2012,Dong2017,Baidya2017,Baars2017}.

An important consideration for boundary layers and other shear flows is the presence of large-scale anisotropy. The spanwise and wall-normal velocity components have narrower power-law scaling regions than the streamwise component \citep{Saddoughi1994}. Further, the assumption of local small-scale isotropy leading to the prediction for inertial subrange scaling \citep{Kolmogorov1941} is often violated for higher-order statistics \citep{Shen2000}. Despite these limitations, the energy spectrum and second-order structure function for the streamwise velocity exhibit power-law behavior consistent with the original predictions of Kolmogorov's theory, even for modest Reynolds numbers \citep[e.g.,][]{Bradshaw1969,Saddoughi1994,Byers2021}. Accordingly, the present work focuses on the geometric features and scaling behavior specific to the streamwise velocity component and its second-order statistics. Attributes of the spanwise and wall-normal velocity statistics are not explored herein.

The remainder of the article is organized as follows: \S \ref{section2} summarizes the measurements and methodology; \S \ref{section3} presents the primary results; and \S \ref{section4} summarizes and discusses the findings in the context of turbulence phenomenology. Detailed accounts of the methodologies are provided in Appendices \ref{appendixA}, \ref{appendixB}, and \ref{appendixC}.

\section{Methodology}
\label{section2}

\subsection{Experiment}

The measurements presented herein were acquired using particle image velocimetry (PIV) in the High Reynolds Number Boundary Layer Wind Tunnel at the University of Melbourne. A brief summary of relevant details is provided here and in table \ref{table1}. A full account of the experiment is given elsewhere \citep{deSilva2014}.

\begin{table}
\begin{center}
\def~{\hphantom{0}}
\begin{tabular}{ccccccl}
$Re_{\tau}$	& $U_\infty$		& $\delta$	& $\nu/u_\tau$	& $L_x/\delta$	& $l^+$	& Source					\\
			& (m\,s$^{-1}$)	& (m)		& ($\mu$m)		& 				&		& 						\\
\hline
12\,300     	& 20				& 0.3	    & 24				& 2				& 37		& \citet{deSilva2014}	\\
\end{tabular}
\caption{Parameters for the particle image velocimetry (PIV) experiment of a smooth-wall boundary layer.}
\label{table1}
\end{center}
\end{table}

The experiment was conducted under approximately zero-pressure-gradient conditions with free stream velocity $U_\infty=$ 20 m\,s$^{-1}$. The boundary layer thickness $\delta=$ 0.3 m is defined using the convention $\delta_{99}=z(U{=}0.99U_\infty)$, where $z$ is the wall-normal position. The friction Reynolds number under these conditions is $\Rey_\tau=\delta u_\tau / \nu =$ 12\,300, where $u_\tau$ is the friction velocity and $\nu$ is the kinematic viscosity of the air in the wind tunnel. For reference, the Taylor microscale Reynolds number is in the range $\Rey_\lambda = 450-550$ across the region of interest.

An eight-camera setup spanning a large field of view was employed to capture PIV measurements in the streamwise--wall-normal ($x{-}z$) plane. The multi-camera setup spatially resolved a wide range of scales between the interrogation window size $l^+=$ 37 and the overall streamwise extent of the field $L_x=2\delta$. Throughout this work, the superscript ``$+$'' indicates wall normalization using $u_\tau$ for velocity and $\nu/u_\tau$ for length. This spatial range captures a majority of the turbulent scales -- excluding the Kolmogorov microscales and very-large-scale motions -- in a high-Reynolds-number setting. Even though the smallest motions are unresolved, the resolution is sufficient to identify the transition from the inertial subrange to the dissipative scales as discussed in \S \ref{section2_3}. The temporal resolution of the PIV is coarse enough such that each PIV velocity field is treated as an independent realization, and ensemble statistics are computed across realizations.

\subsection{Detection of uniform momentum zones}
\label{section2_2}

Relatively uniform flow regions manifest as peaks in histograms of the local streamwise velocity field $u(x,z)$ \citep{Adrian2000}, where the velocity corresponding to each peak is representative of a distinct UMZ \citep{deSilva2016}. At a given instant in time, the number of peaks in the histogram of $u(x,z)$ indicates the number of UMZs in the local flow field. However, this detection is sensitive to the streamwise extent $\mathcal{L}_x$ contributing to each histogram, i.e. how ``local'' the flow field is. In high-Reynolds-number flows, there may be several smaller UMZs across a wide field $\mathcal{L}_x \sim \delta$, where these UMZs collectively smooth the histogram such that the individual peaks are concealed \citep{deSilva2016}. Thus, it is often preferable to employ a narrower field scaled in viscous units $\mathcal{L}_x^+ \sim O(10^3)$ \citep{deSilva2016} or a fraction of the outer length scale $\mathcal{L}_x/\delta \sim O(0.1)$ \citep{Heisel2020}, depending on the region and statistic of interest.

At the same time, the present analysis relies on the continuity of detected UMZ interfaces across the entire 2$\delta$-wide flow field. A multi-step detection method is employed here to account for these seemingly contradictory requirements. The method combines the two most common approaches in the literature: the histogram-based detection of UMZs \citep{deSilva2016} and the fuzzy clustering detection of their interfaces \citep{Fan2019}. In the first step, the PIV field is divided into twelve segments of width $\mathcal{L}_x^+=$ 2\,000. The local histogram is computed for each segment to estimate the number of peaks (i.e. UMZs) in the segment. The average number of UMZs across all segments, rounded to the nearest integer, is assumed to be representative of the 2$\delta$-wide field.

In the second step, the position of UMZ interfaces are detected using fuzzy clustering as proposed in \citet{Fan2019}, where the inputted number of clusters is based on the number of UMZs determined previously. The clustering algorithm assigns each streamwise velocity data point into a cluster such that the overall velocity variance within clusters is minimized. The velocity of the interfaces, corresponding to the boundaries between clusters, is given by the midpoint between cluster centroids. Here, ``boundary'' and ``centroid'' refer to the velocity attributes of the clusters and not their spatial properties. Finally, the position of the UMZ interfaces are determined using isocontours of their velocity. Further details, including an example of the histogram peak detection and a comparison of detected UMZ interfaces with alternate approaches, are provided in Appendix \ref{appendixA}.

\begin{figure}
\centerline{\includegraphics{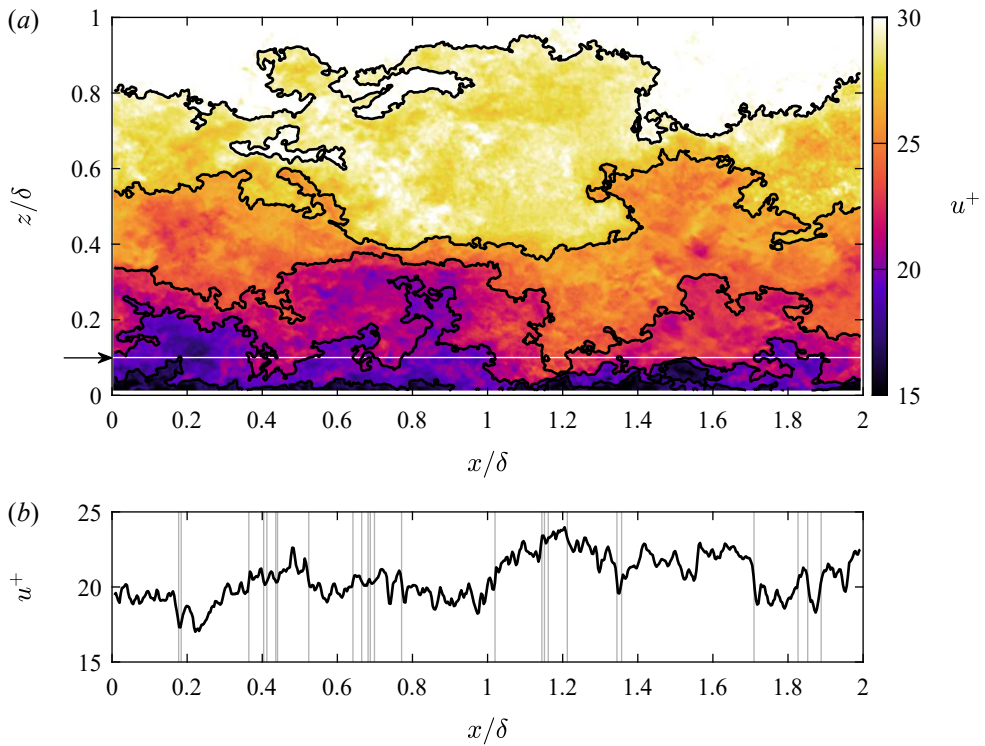}}
\caption{Example of detected streamwise velocity isosurfaces corresponding to the interfaces of uniform momentum zones (UMZs) for a $Re_\tau$ = 12\,300 boundary layer. (\textit{a}) Streamwise velocity $u(x,z)$ in the streamwise--wall-normal measurement plane, overlaid with detected isosurfaces. (\textit{b}) A 1D segment of the velocity signal at the wall-normal position $z = 0.1 \delta$, indicated by the arrow and white horizontal line in (\textit{a}). The vertical lines are the streamwise $x$ positions where the signal intersects the isosurfaces. The same example field is used in all subsequent figures.}
\label{figure1}
\end{figure}

An example flow field with detected UMZ interfaces is shown in figure \ref{figure1}(\textit{a}). Hereafter, we refer to the UMZ interfaces as velocity isosurfaces owing to their definition. While the present planar measurements limit the detection to isolines on a plane rather than surfaces in a volume, the UMZs and their interfaces have been previously observed in three dimensions \citep[e.g.,][]{Chen2020}. The primary limitation of the planar detection employed here is uncertainty in how UMZs connect in the spanwise dimension. For instance, two separate zones of similar momentum may be part of a larger meandering structure \citep{Laskari2018}. Additionally, the detected isosurfaces often delineate isolated UMZ ``pockets'', e.g. near the coordinate ($x{\approx}0.4\delta$, $z{\approx}0.65\delta$) in figure \ref{figure1}(\textit{a}). Three-dimensional measurements are required to determine whether these pockets are connected to a larger UMZ and isosurface in the spanwise direction or are due to small-amplitude velocity variability within a UMZ. To avoid potentially spurious detections in the latter case, small pockets with an enclosed area less than 10$\lambda_T^2$ are removed, where the Taylor microscale $\lambda_T$ is the relevant scaling parameter for the interfaces \citep{Eisma2015,deSilva2017,Heisel2021}. The chosen minimum area ensures the interface thickness encompasses no more than half the area of the pockets. While the pocket near ($x{\approx}0.4\delta$, $z{\approx}0.65\delta$) exceeds the minimum area and is therefore retained, the nearby pockets identified as white-colored regions are smaller than the threshold and are not included. Appendix \ref{appendixB} evaluates how the selected minimum area of pockets affects later results, specifically the fractal dimension of the isosurfaces.

A particular point of interest is the imprint of the detected UMZs and their interfaces on a one-dimensional (1D) velocity signal and its corresponding statistics. Figure \ref{figure1}(\textit{b}) shows an example 1D transect of $u(x,z{=}0.1\delta)$ from panel \ref{figure1}(\textit{a}), where the vertical lines indicate crossings of the isosurfaces. The intermittency of the crossings is immediately apparent. As expected, crossings occur at the largest ``jumps'' in velocity near $x \approx \delta$ and $x \approx 1.7\delta$. Previous studies have shown that regions of high shear generally align with UMZ interfaces \citep{deSilva2017,Gul2020,Chen2021}. However, due to the continuity of the isosurfaces, the interfaces also extend into regions where the shear magnitude is lower. This attribute of the isosurfaces is apparent near $x\approx0.4\delta$ and $x\approx0.7\delta$ in \ref{figure1}(\textit{b}) where the clusters of crossings do not align with large velocity changes. Lastly, the filtering of UMZ pockets is evident near $x \approx 0.25\delta$ and $x \approx 0.75\delta$ where a crossing is excluded because the corresponding spatial region in panel \ref{figure1}(\textit{a}) is too small.

\subsection{Limits of the inertial subrange}
\label{section2_3}

The methods used to estimate the approximate limits of the inertial subrange are discussed here due to variability in conventions across and within disciplines. As discussed in the introduction, the limits are specifically for statistics of the streamwise velocity component. Later figures indicate these subrange limits for the purpose of placing trends and transitions within the spectrum of turbulent scales. The scaling arguments employed here are preferred over directly detecting the --5/3 region in the energy spectrum due to uncertainties in both the spectrum estimate and detection procedure. The estimated limits are supported by later statistics including figure \ref{figure5}(\textit{b}).

The transition from the inertial subrange to the dissipative scales is approximated based on the Kolmogorov lengthscale $\eta = (\nu^3/\epsilon)^{1/4}$, where $\epsilon$ is the average rate of turbulent energy dissipation. The dissipation was estimated as $\epsilon \approx 15 \nu \langle (\partial u / \partial x )^2 \rangle$ assuming local isotropy. Studies have shown the dissipative range to start near the angular wavenumber $k_x = 0.1\eta^{-1}$\citep[e.g.,][]{Saddoughi1994}. The equivalent length in physical space, i.e. $2 \pi / k_x \approx 63 \eta$, is used here as the lower limit of the inertial subrange.

The Taylor microscale $\lambda_T^2 = \langle{u^\prime}^2\rangle / \langle (\partial u / \partial x )^2 \rangle$ has also been used as the upper limit of the dissipative range of scales for a variety of flow applications \citep[see, e.g.,][]{Cava2012,Debue2018,Bandyopadhyay2020}. Further, experimental evidence suggests the intense shear layers and largest vortex cores are proportional to $\lambda_T$ \citep{Eisma2015,deSilva2017,Heisel2021}. Both $\lambda_T$ and $\epsilon$ were estimated here using hotwire anemometry measurements under the same flow conditions. The ratio of the parameters is $\lambda_T/\eta\approx$ 40 to 45 within the region of interest studied herein. Later results include a limited number of data points between the two possible limits $\lambda_T \approx 40\eta$ and $63\eta$, such that there is no meaningful distinction between the two scaling choices with respect to conclusions drawn herein.

The transition from the inertial subrange to the production range is approximated using a dissipation-based lengthscale $L_\epsilon = u_\tau^3/\epsilon$ \citep{Davidson2014}. The friction velocity $u_\tau$ was measured directly from the wall shear stress as $u_\tau = ( \nu \partial U / \partial z ) ^{1/2}$ using separate high-magnification PIV in the viscous sublayer. The measured value was corroborated using a drag balance facility and the Clauser chart method \citep{deSilva2014}. In canonical logarithmic regions with production and dissipation in equilibrium, the length scale simplifies to $L_\epsilon=\kappa z$ and the transition occurs approximately at $z \approx L_\epsilon/\kappa$ \citep{deSilva2015}, where $\kappa$ is the von K\'{a}rm\'{a}n constant. The limit $L_\epsilon/\kappa$ is therefore employed here, noting that the $L_\epsilon$ basis is preferred over $z$ due to its general extension to roughness layers and other non-equilibrium conditions \citep{Davidson2014,Chamecki2017,Ghannam2018}.

With respect to the limits discussed above, the PIV interrogation window size $l/\eta = 14{-}18$ is within the dissipative scales and the streamwise field extent $L_x / L_\epsilon = 20{-}55$ exceeds the inertial subrange. These ranges are based on the $\eta$ and $L_\epsilon$ values observed within the logarithmic region of the boundary layer. The values confirm that the measurement spatial range captures the full extent of the inertial subrange within the region of interest.

\section{Results}
\label{section3}

\subsection{Fractal dimension}
\label{section3_1}

\begin{figure}
\centerline{\includegraphics{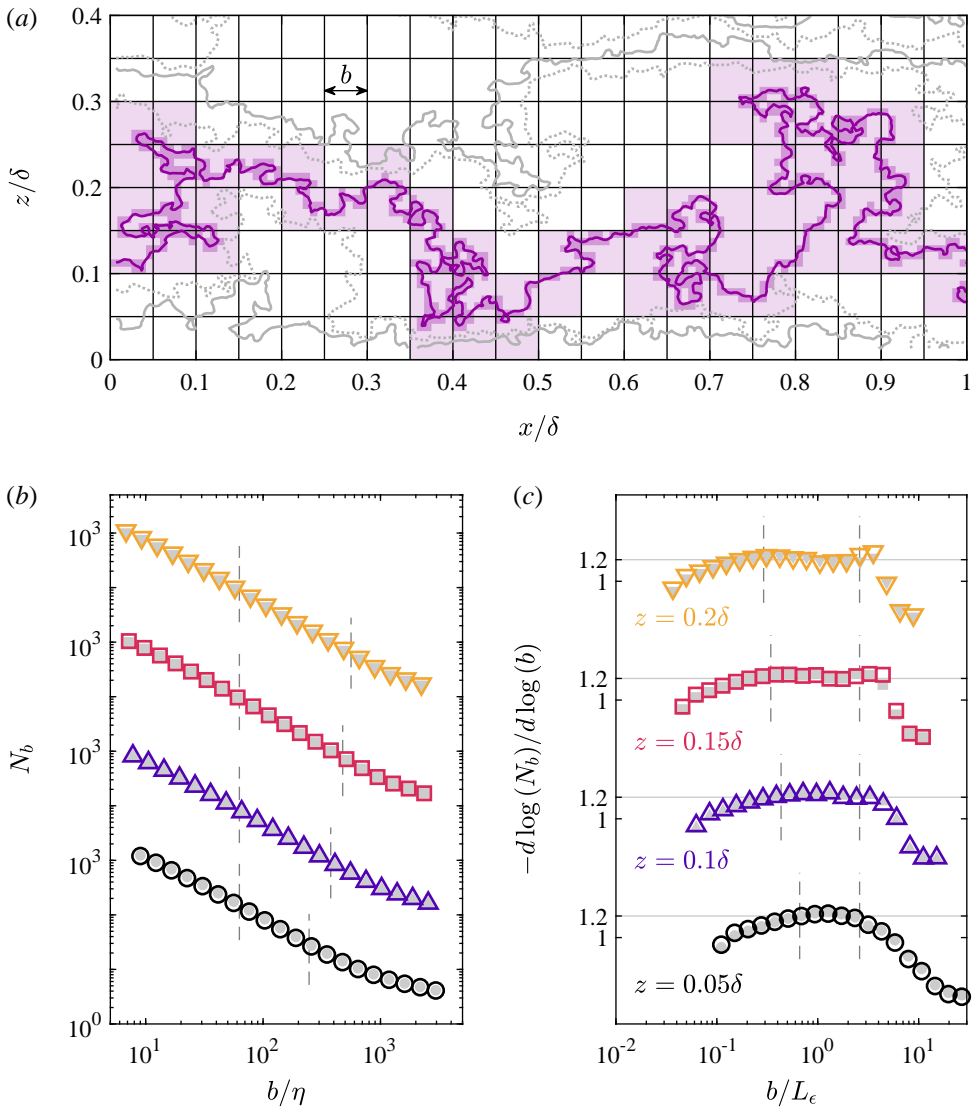}}
\caption{Box counting statistics to estimate the fractal dimension of streamwise velocity isosurfaces. (\textit{a}) An example isosurface (dark line) showing the number of boxes (shaded) of size $b$ required to enclose the isosurface, where the darker shaded boxes are for smaller $b$. Solid lines indicate the UMZ interfaces and dashed lines are internal isosurfaces within the UMZs. (\textit{b}) Number of boxes $N_b$ as a function of box size, where a linear trend indicates statistical self-similarity across scales. (\textit{c}) The local fractal dimension $D_2(b) = - d \log{(N_b)}/d \log{(b)}$ resulting from equation (\ref{equation3_1}). The vertical dashed lines are the approximate limits of the inertial subrange based on the Kolmogorov length scale $\eta$ and the dissipation-based length scale $L_{\epsilon} = u_{\tau}^3/\epsilon$. The statistics are grouped based on the isosurface velocity such that the velocity in each group matches the mean $U(z)$ at $z=0.05\delta$ ($\circ$), $z=0.1\delta$ ($\bigtriangleup$), $z=0.15\delta$ ($\square$), and $z=0.2\delta$ ($\bigtriangledown$). The grey filled symbols are for internal isosurfaces, i.e. the dashed lines in (\textit{a}).}
\label{figure2}
\end{figure}

A direct method for assessing potential fractal geometries is the box counting technique \citep[for applications in turbulence see, e.g.,][]{Sreenivasan1986,Moisy2004,deSilva2013}. In this method, the domain is partitioned into boxes of size $b$, and the number of boxes $N_b$ required to enclose the shape is counted. The process is repeated for a series of box sizes by varying $b$. In principle, the box can have as many dimensions as the domain, e.g. a cube box for a flow volume. A square box in the two-dimensional PIV field is employed here. Figure \ref{figure2}(\textit{a}) shows an example streamwise velocity isosurface and the boxes of size $b$ required to capture the shape.

The statistics for $N_b$ resulting from the box counting routine are shown as colored symbols in figure \ref{figure2}(\textit{b}) for the detected UMZ interfaces. Due to the large range in $z$ spanned by each convoluted isosurface, the statistics are grouped based on the velocity $u_i$ of the isosurfaces. The wall-normal position associated with each isosurface is then taken as the position $z(U{=}u_i)$ where the mean velocity matches $u_i$. The groups in figure \ref{figure2}(\textit{b}) correspond to four $z/\delta$ positions within the logarithmic region. The purpose of associating the velocity groups with a position $z$ is to estimate the inertial subrange limits using local values of $\eta(z)$ and $L_\epsilon(z)$. These limits -- whose definitions are given in \S \ref{section2_3} -- are shown as vertical dashed lines in figure \ref{figure2}(\textit{b}). In figure \ref{figure2} and later figures, the normalizations relevant to both limits of the inertial subrange are shown for completeness.

For statistically self-similar geometries, the resulting relation between $b$ and $N_b$ follows a power law

\begin{equation}
N_b \sim b^{-D_j},
\label{equation3_1}
\end{equation}

\noindent where $D_j$ is the fractal dimension of the geometry and the subscript $j$ is used here to indicate the dimension of the box (e.g. $j=$ 2 for the square box in this case). The shape of $N_b(b)$ reveals the range of scales where the isosurface geometries are self-similar. For instance, the central region of each curve in figure \ref{figure2}(\textit{b}) appears to follow a linear trend in the plotted log-scaled format, consistent with a power law as in equation (\ref{equation3_1}).

A more rigorous test of the power law relation is achieved by estimating the local (in scale) fractal dimension as $D_2(b) = - d \log{(N_b)}/d \log{(b)}$, which is shown in figure \ref{figure2}(\textit{c}). The dimension for each position is approximately constant within the estimated inertial subrange of scales near the value $D_2 \approx 1.2$ represented by horizontal lines in figure \ref{figure2}(\textit{c}). This value matches previous estimates of $D_2$ using the same PIV experiment and a different method to detect UMZ interfaces \citep{deSilva2017}. The results are also consistent with fits to each curve in the inertial subrange which yield values for $D_2$ between 1.2 and 1.22. The range of scales exhibiting the constant $D_2 \approx 1.2$ in figure \ref{figure2}(\textit{c}) is specifically confined to the inertial subrange, which grows with increasing distance from the wall and spans a full decade for $z/\delta=0.2$ ($\bigtriangledown$). Previous studies of UMZ interfaces employed narrower streamwise segments $\mathcal{L}_x = 2\,000 \nu/u_\tau \approx 0.2 \delta$ such that the upper limit of the inertial subrange was not evident in the box-counting results \citep{deSilva2017}. While there is evidence for a constant fractal dimension for scalar concentration isosurfaces in simulations of isotropic turbulence \citep{Iyer2020}, the monofractal behavior confined to the inertial subrange in figure \ref{figure2} is new for velocity isosurfaces within the boundary layer.

Several tests were conducted to assess the robust nature of the results in figure \ref{figure2}. The first test identified velocity isosurfaces between the UMZ interfaces. For a UMZ whose bounding interface isosurfaces are defined by the velocities $u_i$ and $u_{i+1}$, an isosurface of the velocity $0.5(u_i+u_{i+1})$ was assumed to represent facets internal to the UMZ. The dashed lines in figure \ref{figure2}(\textit{a}) correspond to these internal isosurfaces, and the grey filled symbols in figure \ref{figure2}(\textit{b,c}) show the corresponding box counting results. The internal isosurfaces have the same self-similar behavior as the UMZ interfaces, where the fractal dimension $D_2 \approx 1.2$ is constant within the inertial subrange. The second test analyzed isosurfaces of the fixed velocity $U(z{=}0.1\delta)$ which is independent of the UMZ detection. Again, the same result for $D_2$ (not shown here) was observed. The final test repeated the analysis using measurements under varying flow conditions \citep{deSilva2014} including for rough surfaces \citep{Squire2016}, and the same quantitative results were observed. The findings of these tests suggest the constant fractal dimension within the inertial subrange is a general feature of velocity isosurfaces internal to the boundary layer, and is not a unique property of UMZ interfaces dependent on the present flow conditions.

As discussed elsewhere \citep{deSilva2017}, the observed fractal dimension $D_2 \approx 1.2$ is lower than the approximate value 1.33 reported for the turbulent-nonturbulent interface (TNTI) \citep{Sreenivasan1986,deSilva2013,Chauhan2014,Borrell2016}. One-dimensional box counts presented in appendix \ref{appendixC} suggest the lower $D_2$ value estimated here may be due to anisotropy in the shape of the largest geometric features along the velocity isosurfaces. The expected one-dimensional fractal estimate $D_1\approx$ 1/3 is achieved by excluding local regions where the largest features yield a trivial box-counting result. As seen in appendix \ref{appendixB}, the dimension $D_2$ is also sensitive to the treatment of ``pockets'' described in \S \ref{section2_2}. Filtering of pockets introduces a subjective distinction between small-scale features and isosurfaces that should be counted towards statistical self-similarity. Based on appendices \ref{appendixB} and \ref{appendixC}, the differing values for $D_2$ may be due to limitations and subtle differences in methodology. Accordingly, the critical result here for $D_2$ is its constancy within the inertial subrange of scales rather than its precise numerical value.

\subsection{Crossing distributions}
\label{section3_2}

As introduced previously in figure \ref{figure1}(\textit{b}), the convoluted isosurfaces appear as ``crossings'' along 1D streamwise transects of the flow. These transect crossings provide low-order information on the position of velocity changes along $x$. The information is considered low order because the complexity of the continuous velocity signal is reduced to discrete instances where the signal crosses the isosurface velocity $u_i$.

The crossings of detected UMZ interfaces are closely related to so-called zero crossings \citep{Liepmann1949,Liepmann1951,Badri1977,Sreenivasan1983,Kailasnath1993,Poggi2009}. In a zero crossing analysis, point measurements such as hotwire anemometry are used to locate where the velocity signal crosses the mean velocity, i.e. where the fluctuating velocity $u^\prime=u-U$ is zero. In the context of the present analysis, zero crossings are restricted to a single streamwise velocity isosurface $u_i=U(z)$ for a transect measured at $z$. The crossings presented here are generalized to allow for multiple isosurfaces that can differ from the local mean velocity.

\begin{figure}
\centerline{\includegraphics{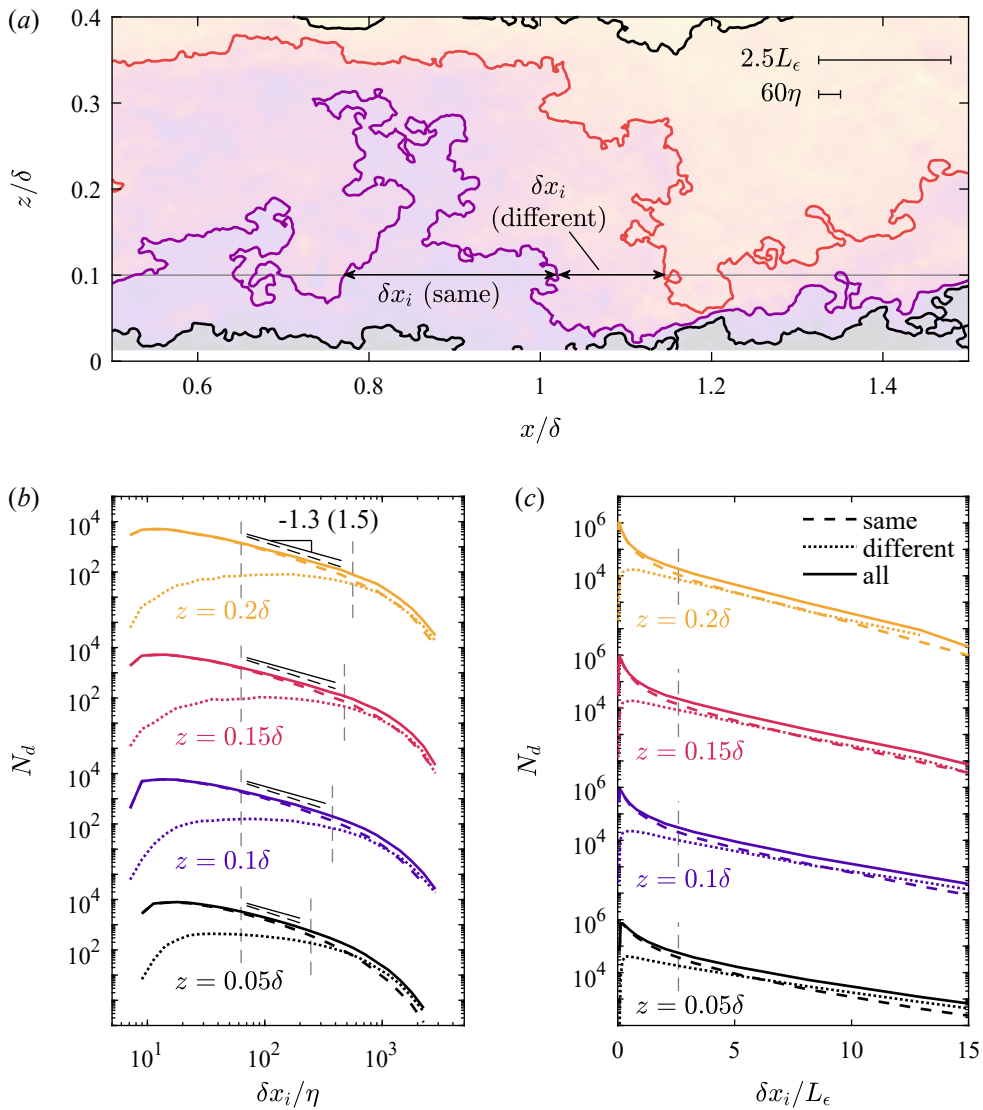}}
\caption{Statistics for the streamwise distance $\delta x_i$ between detected isosurfaces. (\textit{a}) Example isosurfaces showing the distance $\delta x_i$ between crossings of the 1D signal with the same and different isosurfaces. The approximate limits of the inertial subrange for $z=0.1\delta$ are indicated in the upper right corner. (\textit{b}) Distribution of number density $N_d$ for the distances, where the vertical lines are the inertial subrange limits. (\textit{c}) The same results as (\textit{b}) presented in log-linear format.}
\label{figure3}
\end{figure}

An important property of the crossings is the streamwise distance $\delta x_i$ between consecutive isosurface crossings. For reference, the zero crossings literature refers to the distance between crossings as the interpulse period \citep{Sreenivasan2006,Cava2012} or the persistence \citep{Perlekar2011,Chamecki2013,Chowdhuri2020}. Figure \ref{figure3}(\textit{a}) illustrates the definition of $\delta x_i$ using isosurfaces of detected UMZ interfaces that cross a 1D signal at $z=0.1\delta$. Because multiple UMZ interfaces are present within the boundary layer, the segments of length $\delta x_i$ can be bounded by either the same velocity isosurface or different isosurfaces. Separate statistics are presented for these two scenarios.

Figure \ref{figure3}(\textit{b,c}) shows probability distributions of $\delta x_i$ for transects along four wall-normal positions within the log region. The distributions are presented as number densities $N_d$, where the total area under the distribution represents the number of measured $\delta x_i$ events. The number density allows for a direct comparison between the two crossing scenarios described above: for a given value of $\delta x_i$, the scenario with larger $N_d(\delta x_i)$ is relatively more frequent and thus contributes more to overall statistics. The distribution for all values of $\delta x_i$, shown as a solid line figure \ref{figure3}(\textit{b,c}), is simply the sum of the individual $N_d$ distributions.

The distribution shape for $\delta x_i$ between crossings of the same isosurface is in close agreement with previous observations for the zero crossing interpulse period. Intermediate distances are well approximated by a power law \citep{Bershadskii2004}, and the shape transitions to an exponential tail across longer distances \citep{Sreenivasan1983,Cava2012,Chamecki2013}. The exponential trend is more apparent from the log-linear plots presented in figure \ref{figure3}(\textit{c}). Previous studies have described the shortest interpulse periods using a log-normal distribution \citep{Badri1977,Sreenivasan2006}, but the present PIV measurements do not resolve the trends at the dissipative scales where the log-normal distribution is expected.

Within the inertial subrange of scales, the power law exponent for $N_d(\delta x_i)$ between crossings of the same isosurface is approximately --1.5 regardless of position $z$. The slope again agrees with results for zero crossings \citep{Bershadskii2004,Cava2012}. Based on the amplitude of $N_d$ in figure \ref{figure3}(\textit{b}), $\delta x_i$ values corresponding to the inertial subrange are predominately due to crossings of  the same isosurface. Even though $\delta x_i$ values between different isosurfaces are not self-similar in this region, the behavior along a single isosurface has leading-order importance such that the overall curve still approximates a power law. However, the overall power law exponent (--1.3) is somewhat flattened by the contributions of $\delta x_i$ between crossings of different isosurfaces. The trends in figures \ref{figure2} and \ref{figure3}(\textit{b}) demonstrate that the signature of self-similarity in the inertial subrange is reflected by the geometry of individual streamwise velocity isosurfaces. The behavior across consecutive isosurfaces, which is related to the amplitude of velocity variations, affects the power law exponent but does not contribute directly to the self-similarity.

Across longer $\delta x_i$ crossing intervals, the $N_d$ trends transition and the statistics between different isosurfaces gain leading-order importance within the production range of scales. The exact transition point where the $N_d$ curves intersect is sensitive to the detection of UMZs, and the result is discussed here qualitatively. The transition corresponds to a shift from a power law distribution to an exponential tail seen in figure \ref{figure3}(\textit{c}). The tail slope varies moderately between wall-normal positions, suggesting a normalization parameter other than $L_\epsilon$ is required to describe variations in the tail slope. The results agree with previous works that observed the exponential tail to deviate from the integral scales \citep{Chamecki2013}. This trend is not the focus of the present analysis, and is not explored further here.

The results in figure \ref{figure3} were reproduced using randomized data sets to ensure the $N_d$ distributions are not an artifact of the methodology. Synthetic velocity fields generated using random Gaussian noise do not contain any large-scale organization or UMZ signature \citep{deSilva2016}, and all observed trends in $N_d$ are lost. In a separate test, the streamwise velocity fluctuations in each PIV field were randomized in their phase \citep{Theiler1992}. Phase randomization of the two-dimensional Fourier transform preserves the original two-dimensional energy spectrum and correlations of the velocity. The phase-randomized data produces the same overall shape of $N_d$ due to the relation between the crossing distributions and the spectrum \citep{Bershadskii2004,Poggi2009,Heisel2022}. However, there are fewer crossings between different isosurfaces and its distribution is no longer exponential within the inertial subrange. The result indicates that the observed transition in statistics between the inertial and production ranges is related to aspects of the flow structure that are distorted during phase randomization.

The figure \ref{figure3} trends can be used to interpret the transition from the inertial subrange to the production range of scales in the context of coherent structures. The $\delta x_i$ values between different isosurfaces represent the distances between adjacent UMZ edges. The distance across UMZs, and more generally the overall geometry of the large-scale velocity regions, becomes the limiting factor within the production range. The confinement of the self-similar isosurfaces by the large-scale organization of the flow is consistent with the sharp decline of $N_d$ in figure \ref{figure3}(\textit{b}) and the deviation from a constant fractal dimension in figure \ref{figure2}(\textit{c}). In this sense, the statistically relevant property of the flow geometry is the self-similarity of isosurfaces in the inertial subrange, and the size of the coherent velocity regions in the production range. This interpretation of the production range is essentially the same as the existing viewpoint \citep{Perry1982} and is supported by direct observations of UMZ sizes \citep{Heisel2018,Heisel2020} and distances between coherent velocity regions \citep{Dong2017}.

\subsection{Conditional structure functions}

The remainder of the analysis quantifies how the self-similar geometries influence scale-dependent statistics such as the structure function. The second-order longitudinal structure function of the streamwise velocity is defined as

\begin{equation}
\langle \Delta u^2 \rangle (r_x) = \langle \left[ u(x_o+r_x)-u(x_o) \right]^2 \rangle,
\label{equation3_2}
\end{equation}

\noindent where $x_o$ is the reference point for each instantaneous structure function, $r_x$ is the streamwise distance from the reference, and angled brackets ``$\langle \cdot \rangle$'' indicate an ensemble average across all $x_o$ at a given wall-normal position. The function quantifies the cumulative velocity increment from the reference point up to distance $r_x$. Only the second-order longitudinal structure function is discussed in this study, and hereafter $\langle \Delta u^2 \rangle$ is referred to as ``the structure function'' for simplicity.

\begin{figure}
\centerline{\includegraphics{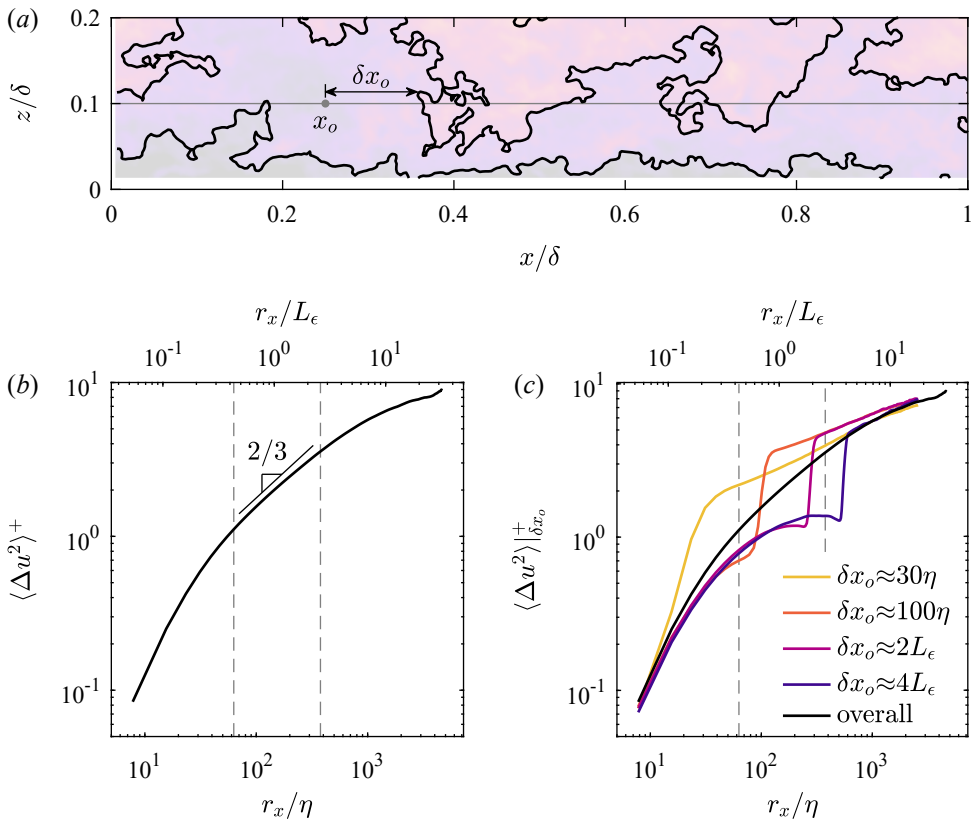}}
\caption{Second-order structure function statistics conditionally averaged on the proximity to the detected isosurfaces at $z=0.1\delta$. (\textit{a}) Example isosurfaces showing the distance $\delta x_o$ to the nearest crossing of an isosurface with the 1D signal, where $x_o$ is the reference point for an instantaneous velocity increment $\Delta u(x_o,r_x) = u(x_o+r_x)-u(x_o)$. (\textit{b}) Structure function  $\langle \Delta u^2 \rangle (r_x)$ ensemble-averaged across all $x_o$, where the vertical lines are the approximate limits of the inertial subrange. (\textit{c}) Structure functions conditionally averaged across instances of $x_o$ where the distance $\delta x_o$ matches the value specified in the legend.}
\label{figure4}
\end{figure}

To assess the contribution of spatial features such as the detected streamwise velocity isosurfaces to the structure function statistics, the ensemble averaging operator in equation (\ref{equation3_2}) can be replaced by a conditional averaging operator based on an additional parameter related to the spatial features. The parameter $\delta x_o = x_i-x_o$ describes the distance from a given reference point $x_o$ to the nearest crossing $x_i$ of a detected streamwise velocity isosurface. Figure \ref{figure4}(\textit{a}) shows the definition of $\delta x_o$ for an example isosurface. Rather than ensemble averaging across all $x_o$, the conditional averaging operator only considers instances where $\delta x_o$ is a specified value. The result is a structure function dependent on both $r_x$ and the distance $\delta x_o$ to the nearest isosurface:

\begin{equation}
\langle \Delta u^2 \rangle (r_x) \vert_{\delta x_o} = \langle \Delta u(r_x \, \vert \, x_o{=}x_i{-}\delta x_o )^2 \rangle,
\label{equation3_3}
\end{equation}

\noindent where the notation ``$\vert_{\delta x_o}$'' is used to indicate the conditional averaging.

The traditional structure function using equation (\ref{equation3_2}) is discussed here prior to introducing the conditional results. Figure \ref{figure4}(\textit{b}) shows $\langle \Delta u^2 \rangle$ for the wall-normal position $z=0.1\delta$. A power law slope of 2 describes the smallest $r_x$ increments, but is not assessed here due to spatial resolution limitations. This power law exponent is predicted from the K\'{a}rm\'{a}n--Howarth equation when viscous diffusion dominates over inertial mechanisms \citep{Pope2000}. In the inertial subrange, Kolmogorov's second similarity hypothesis predicts $\langle \Delta u^2 \rangle \sim r_x^{2/3}$, which is well-supported by measurements and simulations \citep{Pope2000}. However, the 2/3 signature in figure \ref{figure4}(\textit{b}) is not strictly constant within the inertial subrange. The changing slope within the inertial subrange is a known limitation of structure function statistics, and is in part due to the inclusion of other small- and large-scale effects in the cumulative velocity increment $\Delta u$ \citep[see, e.g.,][]{Davidson2005}.

Conditional structure functions $\langle \Delta u^2 \rangle \vert_{\delta x_o}$ for four values of $\delta x_o$ are shown in figure \ref{figure4}(\textit{c}). The four values were chosen to represent a range within and beyond the inertial subrange. By fixing the position of the nearest detected isosurface $\delta x_o$, and considering the isosurfaces are a low-order representation of velocity changes as discussed previously, it is expected that the resulting conditional average will yield a large velocity increment near $r_x \approx \delta x_o$. Large ``jumps'' in $\langle \Delta u^2 \rangle \vert_{\delta x_o}$ are indeed observed at $r_x \approx \delta x_o$. Further, each conditional curve converges to the overall result $\langle \Delta u^2 \rangle$ across long distances where the imposed $\delta x_o$ condition is no longer relevant. The thickness of the jumps (in terms of $r_x$) is in part due to viscous effects that act to smooth the velocity gradients. Accordingly, the jump thickness for $\delta x_o \approx 30 \eta$ spans most of the dissipative range, but appears increasingly sharp for the other cases along the logarithmic abscissa.

The behavior of the jumps in $\langle \Delta u^2 \rangle \vert_{\delta x_o}$, in comparison to the approximate power law within the inertial subrange, demonstrates the features underlying the ensemble-averaged structure function. The power law relation is not apparent from any given instantaneous velocity increment $\Delta u(x_o,r_x) = u(x_o+r_x)-u(x_o)$. Rather, it is the stochastic result of variability in the positions of the isosurfaces across numerous instances, where the position is parameterized as $\delta x_o$ in figure \ref{figure4}. The distributions in figure \ref{figure3} already showed that the variability is statistically self-similar in the inertial subrange, a direct result of the fractal dimension in figure \ref{figure2}. The structure function power law may therefore result from a composite of ``jumps'' across self-similar velocity isosurfaces. This possibility is explored in the following section using a simplified velocity signal.

\subsection{Simplified velocity signal}

The crossing positions in \S \ref{section3_2} (figure \ref{figure3}) encompass the geometric properties of the isosurfaces evaluated in \S \ref{section3_1} (figure \ref{figure2}). Here, the measured 1D velocity signal is reduced to these crossing properties. By removing aspects of the signal unrelated to the isosurface crossings, the simplified signal is used to demonstrates how the self-similar signature of the multi-dimensional isosurface geometries translates to scale-dependent statistics of the longitudinal signal, specifically the structure function.

In a zero crossing analysis, the simplification is accomplished using the telegraph approximation (TA) \citep[e.g.,][]{Bershadskii2004}. The TA signal is assigned a value of 1 or 0 depending on the sign of the fluctuating velocity: $u_{TA}=1$ where $u^\prime>0$ and $u_{TA}=0$ where $u^\prime<0$. The basic principles of the TA signal are generalized here to construct a simplified signal $u_\pm$ based on the detected streamwise velocity isosurfaces. Hereafter, $u_\pm$ is referred to as the ``iso-crossing'' signal.

The 1D iso-crossing signal is assigned an initial value $u_\pm(x{=}0)=0$. At each isosurface crossing position $x_i$, $u_\pm$ is increased by 1 if $\partial u / \partial x > 0$ and is decreased by 1 if $\partial u / \partial x < 0$ based on the gradient at the crossing. The iso-crossing signal therefore contains information on the position and sign of the velocity change for each crossing. The signal is equivalent to $u_{TA}$ for mean velocity crossings $u_i=U$, and additionally allows for multiple isosurfaces of any arbitrary velocity $u_i$.

\begin{figure}
\centerline{\includegraphics{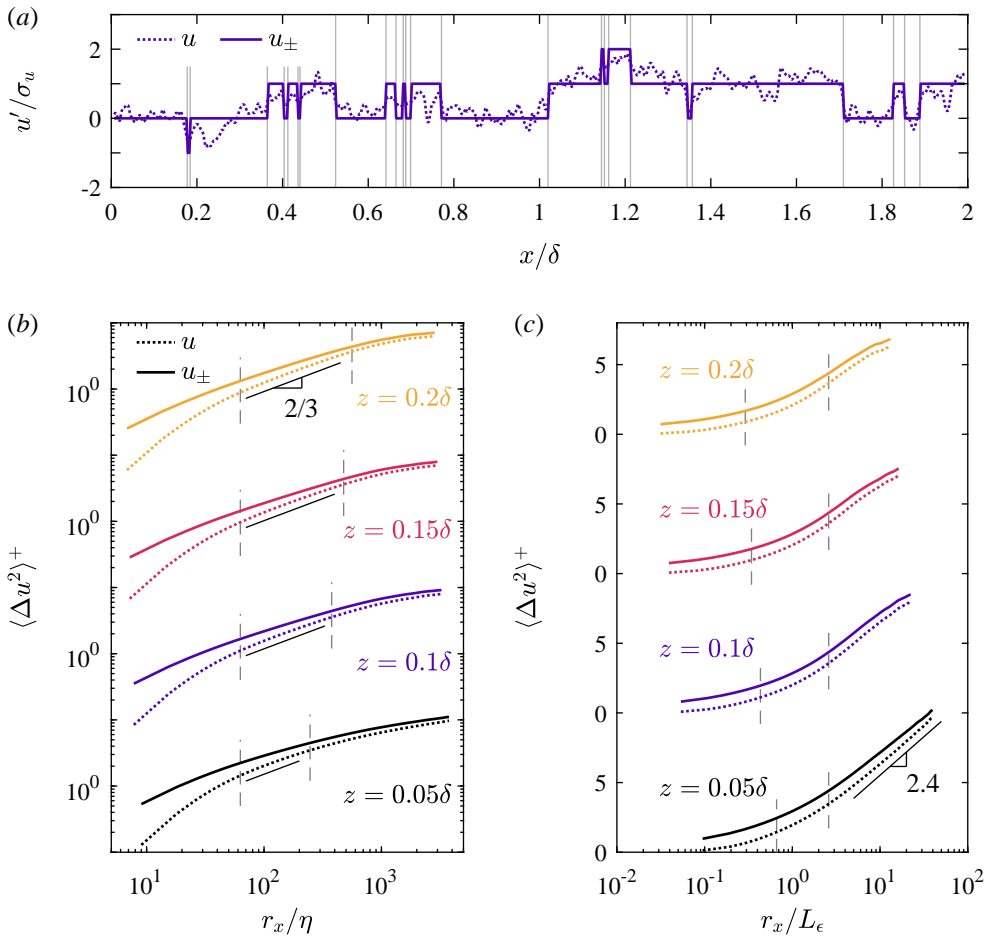}}
\caption{Statistics for a simplified velocity signal $u_\pm$. (\textit{a}) Example 1D velocity signal, where the simplified signal $u_\pm$ retains only the position of isosurface crossings (indicated by vertical lines) and the sign of the velocity difference at the crossing. (\textit{b}) Second-order structure functions of the measured $u$ and simplified $u_\pm$ signals, where the vertical lines are the approximate limits of the inertial subrange and the results for $u_\pm$ are shifted for comparison. (\textit{c}) The same results as (\textit{b}) presented in log-linear format.}
\label{figure5}
\end{figure}

Figure \ref{figure5}(\textit{a}) shows an example iso-crossing signal in comparison with the measured velocity $u$. To negate the effect of the velocity magnitude in the visual comparison, the $u$ signal is normalized to achieve the same standard deviation as the $u_\pm$ signal, and the values are shifted to achieve $u(x{=}0)=0$. The presence of multiple isosurfaces with different velocities $u_i$ leads to $u_\pm$ values beyond 0 and 1. The four observed values of $u_\pm$ ranging from --1 to 2 indicate that the example 1D signal spans four UMZs.

Structure function statistics for $u$ and $u_\pm$ are presented in figure \ref{figure5}(\textit{b,c}). To facilitate comparison of the trends, the iso-crossing structure functions are increased by a constant factor such that the curves are shifted closer to the measured results. The shift does not affect the shape or slope of the curves in figure \ref{figure5}(\textit{b}). The shift does increase the slope of the linear region for the $u_\pm$ curves in figure \ref{figure5}(\textit{c}), but it does not change the fact that an approximately linear region exists.

The primary difference between the structure functions is the excess variability for the iso-crossing signal at small $r_x$ distances corresponding to the dissipative scales. The larger values are due to the removal of viscous smoothing effects in favor of discontinuous jumps in $u_\pm$. Otherwise, despite the limited information in the iso-crossing signal as seen in figure \ref{figure5}(\textit{a}), the resulting structure function captures several key features of $\langle \Delta u ^2 \rangle$. For figure \ref{figure5}(\textit{b}) in the inertial subrange, a power law is observed with a cleaner signature for $u_\pm$ than $u$. The fitted power law exponent is approximately 0.55 for the iso-crossing signal, which is smaller than the value close to 2/3 observed for $u$. The difference in power law exponent is further discussed later in this section.

For figure \ref{figure5}(\textit{c}) in the production range, the structure functions nearest the wall are approximately linear. The linear trend is consistent with the predicted logarithmic dependence $\langle \Delta u ^2 \rangle \sim \ln(r_x)$ that is analogous to the $k_x^{-1}$ signature in the energy spectrum \citep{Davidson2006,Davidson2006a}. Further, the linear slope for the measured $u$ signal closely matches previous observations of 2.4 \citep{deSilva2015}. However, the linear slope for the iso-crossing structure function is not representative due to the manual shift discussed above.

In addition to capturing the inertial subrange power law and production range logarithmic trends, the iso-crossing structure function also identifies where the transition occurs between the two scaling regions. The result is entirely consistent with the crossing distribution in figure \ref{figure3}. The $N_d$ distributions exhibit an inertial subrange power law owing to the self-similar geometry of the isosurfaces and a transition to an exponential tail at scales where the overall geometry of the large-scale velocity regions gains leading-order importance. Figure \ref{figure5} demonstrates how these attributes of the isosurface geometry and their crossing properties propagate to structure function statistics. The same trends in the transition to the production range are also apparent from the spectrum of a TA signal \citep{Huang2021}, likely due to the TA signal exhibiting the same exponential cutoff in figure \ref{figure3} imposed by the larger-scale geometries. 

\begin{figure}
\centerline{\includegraphics{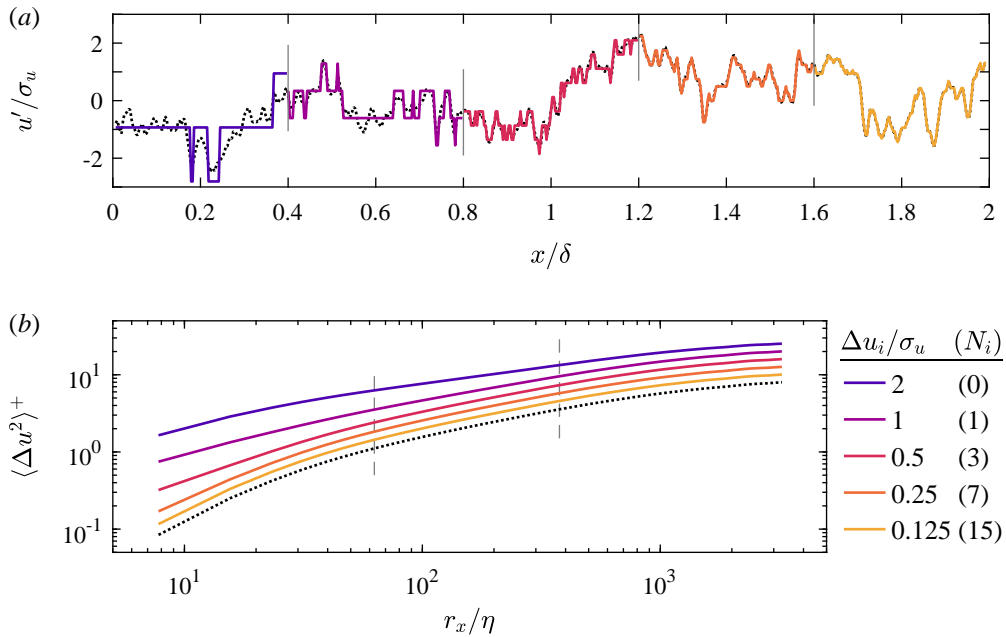}}
\caption{Comparison of the measured velocity $u$ (dotted line) at $z=0.1\delta$ with the iso-crossing signal $u_\pm$ (colored lines) defined using a varying number of isosurfaces. (\textit{a}) Example 1D signal, where the number of isosurfaces given by the line color changes for each $0.2\delta$-wide interval delineated by vertical lines. (\textit{b}) Second-order structure functions, where the vertical lines are the approximate limits of the inertial subrange and the results for $u_\pm$ are shifted for comparison. Here, $N_i$ is the number of new isosurfaces added between detected UMZ interfaces. As the number of isosurfaces increases, the velocity difference $\Delta u_i$ between adjacent isosurfaces decreases and the $u_\pm$ signal more accurately represents $u$.}
\label{figure6}
\end{figure}

The primary attribute absent from the iso-crossing signal is the variability in velocity amplitude at positions between the crossings. This absence is responsible for the difference in power law exponent between the structure functions for $u$ and $u_\pm$ seen in figure \ref{figure5}(\textit{b}). A more representative iso-crossing signal can be achieved by simply incorporating additional isosurfaces at velocity increments between the detected UMZ interfaces. Figure \ref{figure6} uses results at $z=0.1\delta$ to demonstrate how $u_\pm$ and its structure function change with the inclusion of additional isosurfaces. Unlike for previous results, the smallest isolated isosurface pockets are not excluded from the signal in figure \ref{figure6}. While these pockets obfuscate the signature of fractal self-similarity as discussed in \S \ref{section3_1}, the pockets must be included for a full representation of the velocity signal.

The number of isosurface level sets is parameterized by the resulting velocity difference $\Delta u_i$ between adjacent isosurfaces. The iso-crossing signal can be considered as a transformation from spatial coordinates to a velocity grid. Rather than discretizing the signal at fixed positions or times, the signal is mapped to fixed velocities whose resolution is given by $\Delta u_i$. As the number of level sets increases, the iso-crossing signal captures the velocity variability in finer increments of velocity resolution and more closely resembles $u$. The result for $\Delta u_i / \sigma_u=0.5$, shown for $x/\delta=$ 0.8 to 1.2 in figure \ref{figure6}(\textit{a}), features 8 isosurfaces across the range of $u$. Yet the structure function resulting from this $u_\pm$ (ensemble-averaged across all PIV fields) shown in figure \ref{figure6}(\textit{b}) has a power law exponent 0.62, within 10\% of 2/3. An even larger improvement occurs within the dissipative range, where the sharp velocity jumps are distributed across additional steps spanning a wider distance as $\Delta u_i$ decreases.

The additional level sets in figure \ref{figure6} correspond to internal isosurfaces within the detected UMZs. Figure \ref{figure2} showed that the internal isosurfaces exhibit the same self-similarity as the UMZ interfaces. As a result, every streamwise velocity isosurface $u_i$ represented by the $u_\pm$ signals in figure \ref{figure6} will have self-similar crossing distributions analogous to the power law in figure \ref{figure3}. The velocity amplitude variability captured by the finer $\Delta u_i$ increments is therefore expected to also exhibit self-similar behavior, specifically in terms of the distances between positions where the velocity reaches the same amplitude. In this sense, the self-similar signature of individual isosurfaces extends to velocity amplitude upon considering the collective contribution of many isosurfaces to a 1D transect of the flow geometry. While an infinite number of isosurface level sets is required to fully represent the continuous velocity signal, trends across the turbulence spectrum can be reasonably recovered with a limited account of isosurface properties. Most notably, the original iso-crossing signal in figure \ref{figure5} with a minimal number of isosurfaces is able to reproduce a majority of the scale-dependent behavior in the inertial and production regions.

\section{Discussion}
\label{section4}

\begin{figure}
\centerline{\includegraphics{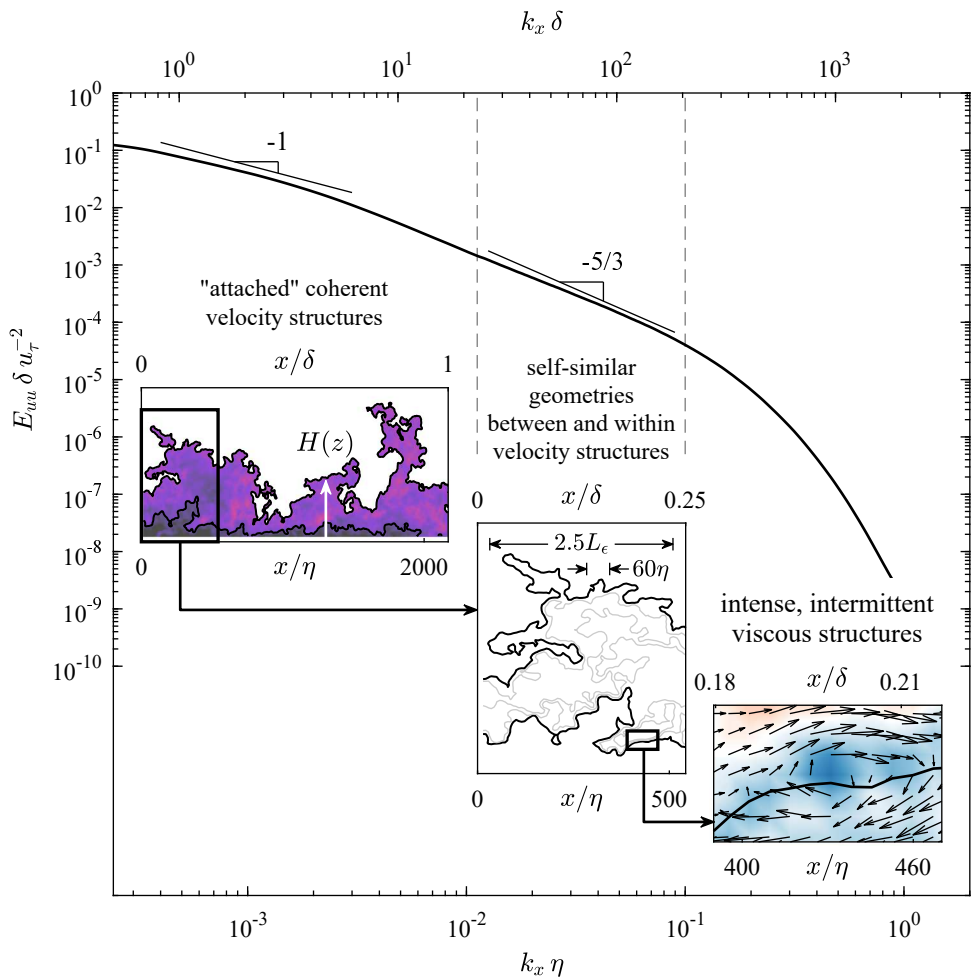}}
\caption{Energy spectrum $E_{uu}$ of the streamwise velocity. The insets show examples of the prominent geometric features corresponding to streamwise velocity statistics in the production range, inertial subrange, and dissipative range (from left to right) of turbulent scales. The color corresponds to streamwise velocity in the left inset and vorticity in the right inset. The spectrum is based on hotwire anemometry measurements under the same flow conditions as the PIV experiment.}
\label{figure7}
\end{figure}

The shape of the largest- and smallest-scale coherent spatial features is well-documented for boundary layer turbulence as noted in the introduction. However, the features are isolated in the sense that there is limited evidence detailing the pathway of intermediate eddies linking these disconnected scales. \citet{Marusic2019} specifically noted the need to better understand the relationship between coherent structures in the production range and eddies in the inertial subrange. While the full complexity of inertial subrange eddies remains inadequately understood, the present evidence characterizes their signature on the geometric structure of the streamwise velocity field. This characterization is illustrated in figure \ref{figure7}.

The signature of the production range is $E_{uu} \sim k_x^{-1}$ in the energy spectrum \citep{Perry1982} and $\langle \Delta u ^2 \rangle \sim \ln(r_x)$ in the structure function \citep{Davidson2006}. This signature corresponds to ``attached'' eddies whose size scales with wall-normal distance \citep{Townsend1976}. These eddies are realized in space as regions of coherent streamwise velocity such as low- and high-speed streaky structures \citep{Hwang2015,Hwang2018} or more generally as UMZs \citep{Heisel2020} as depicted in the left inset of figure \ref{figure7}. The boundaries of these structures are detected here as UMZ interfaces which are often aligned with internal shear layers \citep{Gul2020,Heisel2021}.

Based on the box-counting results in figure \ref{figure2}, the boundaries of these large-scale velocity regions are geometrically self-similar. The boundaries contain ``wrinkles'' of various sizes within the inertial subrange of scales as shown in the center inset of figure \ref{figure7}. The self-similarity of the geometry directly influences statistics of 1D velocity signals, leading to power law signatures in the isosurface crossing properties (figure \ref{figure3}) and structure function statistics (figures \ref{figure4} and \ref{figure5}). The self-similar geometry is observed for streamwise velocity isosurfaces throughout the flow volume, i.e. both along the boundaries of and within the velocity regions. The collective behavior of multiple adjacent isosurfaces is related to variability in the velocity amplitude between the UMZ interfaces (figure \ref{figure6}).

The consistent fractal dimension observed for every tested streamwise velocity isosurface indicates that the self-similar isosurface wrinkles are space-filling. The same is true for the coherent velocity regions that approximately fill the boundary layer outer region \citep{Heisel2020}. This is in contrast to the intermittent small-scale geometries such as tubes and filaments. While weak Kolmogorov-scaled eddies likely exist throughout the flow volume, the most intense and statistically relevant small-scale features tend to cluster along ``active'' regions where the shear and dissipation magnitudes are the largest \citep{She1990,Moisy2004,Ishihara2009}. An example high-vorticity region is shown in the right inset of figure \ref{figure7}.

As seen in figure \ref{figure7} and in previous figures, the isosurface wrinkles appear to span a range of sizes extending to scales both larger and smaller than the approximate limits of the inertial subrange. The inertial subrange is the specific region where the statistical self-similarity is fixed, i.e. the fractal dimension is constant in figure \ref{figure2}. For the larger wrinkles, the departure from self-similarity can be explained by the increased relevance of the  large-scale geometries related to the flow configuration (figure \ref{figure3}). Likewise, the shape of the smallest wrinkles is expected to be influenced by viscous processes such as diffusion. These observations are consistent with theoretical arguments that the inertial subrange must satisfy $\eta \ll r_x \ll L$. We therefore postulate that the inertial subrange eddies are reflected by the wrinkled geometry of streamwise velocity isosurfaces in the range of sizes where no flow length scale has leading-order importance. The self-similarity of these geometries yields monofractal behavior in crossing properties of the velocity.

While not explored in detail here, there are expressions that directly relate the fractal dimension $D_1 \approx 0.33$ (appendix \ref{appendixC}), the power law exponent $\gamma \approx 1.3$ of crossing intervals (figure \ref{figure3}), and the iso-crossing structure function exponent $n \approx 0.55$ (figure \ref{figure5}). For a stochastic process defined by a power law probability distribution with exponent $\gamma$, the statistics are related as $D_1=1-n$ and $n=2-\gamma$ \citep{Heisel2022}. In these relations, the energy spectrum exponent is replaced by the equivalent value $n+1$, and $n=2-\gamma$ is consistent with predictions for a superposition of Poisson processes \citep{Jensen1998}. The expressions are specific for binary stochastic processes like the TA signal, and the relations depend on the phase of the signal if amplitude variability is introduced \citep{Higuchi1990}. In additional to the variable amplitude of the iso-crossing signal, the difference between the predicted values and the iso-crossing values observed here are in part due to the finite power law region \citep{Heisel2022}, which spans approximately one order of magnitude for the given $\Rey_\lambda$. Considering the limitations in the Reynolds number achieved for modern laboratory and numerical experiments, the deviations in the statistical relations caused by a limited range of self-similarity presents a formidable challenge for investigating the role of fractality in turbulence.

Lastly, the present measurements do not allow any characterization of how the isosurfaces dynamically evolve. For instance, it is not known how the smaller and larger wrinkles interact through the dynamics of the Navier-Stokes equations. Previous works have proposed mechanisms including vortex stretching \citep{Taylor1937} and rapid distortion \citep{Hunt2014}. Recent evidence suggests that strain self-amplification \citep{Tsinober2001} is the foremost contributor to energy transfer dynamics in the inertial subrange \citep{Carbone2020,Johnson2020}. Yet, forward energy transfer from large to small scales is only true in the average sense, as inverse transfer or ``backscatter'' is prominent in instantaneous realizations \citep[see, e.g.,][and references therein]{Pope2000,Cardesa2017,Alves2017,Carter2018}. In the context of the velocity isosurfaces, backscatter corresponds to the alignment of large velocity gradients with the smallest wrinkles and relatively smaller velocity differences across the extent of larger wrinkles. Further, randomizing the phase of a TA signal does not change the resulting spectral power law exponent in the inertial subrange \citep{Poggi2009}. The presence of backscatter and noninfluence of phase underscore the point that the inertial subrange power law signature and it exponents are a statistical result achieved through spatial and temporal averaging of the governing physics, where the exponent values may not be representative of local energy transfer. The fact that the self-similarity of the isosurfaces is only fixed in a statistical sense serves to further emphasize this point.

The authors gratefully acknowledge funding from the US National Science Foundation. M.H. is supported by grant NSF-AGS-2031312, and G.K. is supported by grant NSF-AGS-2028633.

Declaration of Interests. The authors report no conflict of interest.

\appendix

\section{Further details on the detection of uniform momentum zones}
\label{appendixA}

The appended materials supplement the methodology in \S \ref{section2_2} with additional details for completeness. As discussed in \S \ref{section2_2}, the number of UMZs is first estimated using histograms on streamwise segments of width $\mathcal{L}_x^+=$ 2\,000, then fuzzy clustering is employed to determine the velocity of the UMZ interface isosurfaces. Details pertaining to both of these steps are provided here.

\begin{figure}
\centerline{\includegraphics{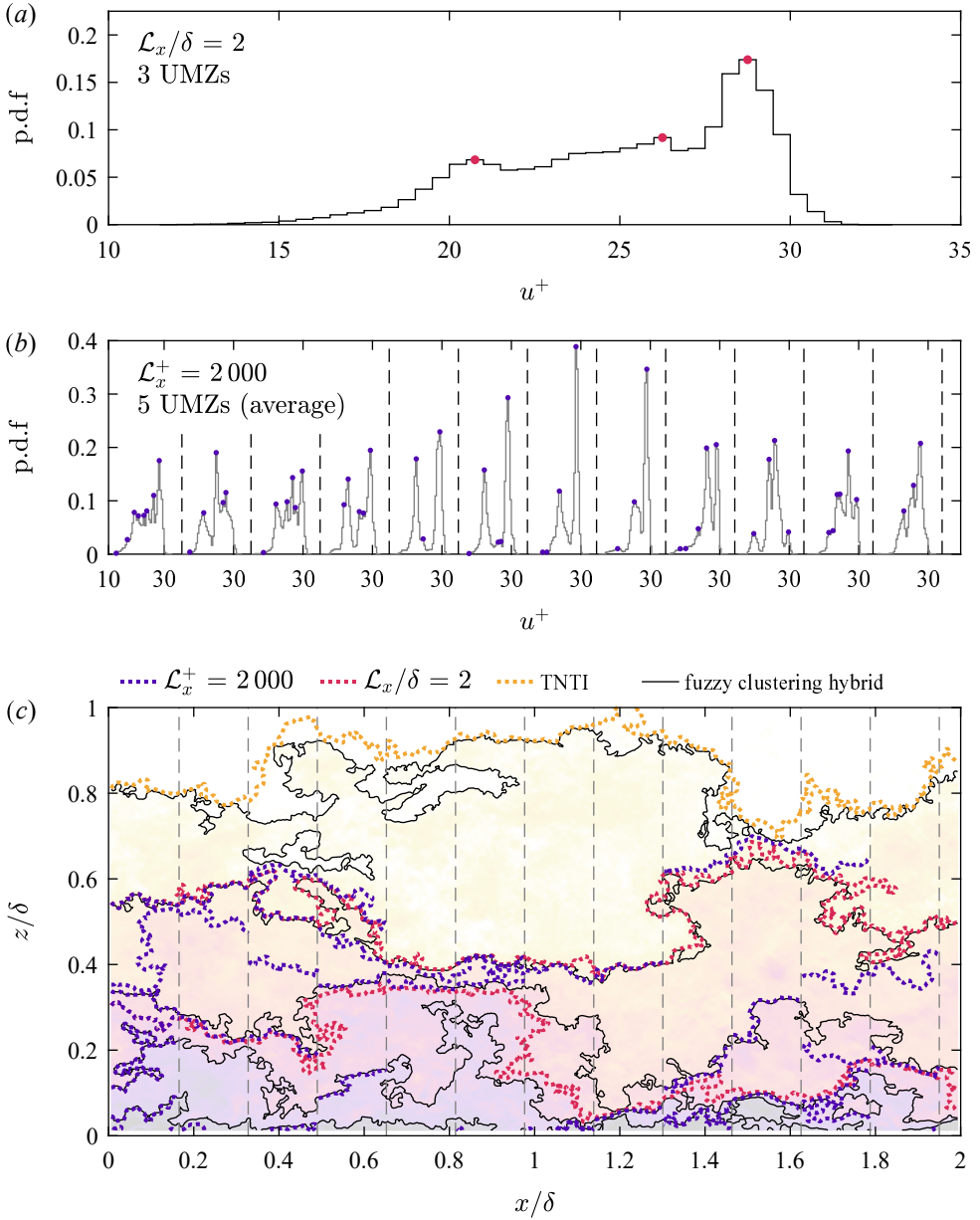}}
\caption{Comparison of detected isosurfaces for the histogram and fuzzy clustering methodologies. (\textit{a}) Histogram of streamwise velocities for the entire PIV field whose length is $\mathcal{L}_x/\delta=$ 2. (\textit{b}) Histograms for sections of length $\mathcal{L}_x^+ =$ 2\,000 within the PIV field. (\textit{c}) Detected isosurfaces for the two $\mathcal{L}_x$ values, the detected turbulent-nonturbulent interface (TNTI), and fuzzy clustering isosurfaces. The inputted number of UMZs for the fuzzy clustering detection is the average number of peaks in (\textit{b}).}
\label{figure8}
\end{figure}

An example of velocity histograms using 0.5$u_\tau$-wide bins is given in figure \ref{figure8}. A histogram based on the entire field width $\mathcal{L}_x/\delta=$ 2 is shown in panel \ref{figure8}(\textit{a}), and the histograms for $\mathcal{L}_x^+=$ 2\,000 ($\mathcal{L}_x/\delta=$0.16) segments are shown in \ref{figure8}(\textit{b}). Velocity values within the free stream region above the turbulent-nonturbulent interface (TNTI) are excluded from the histograms \citep{deSilva2016}. To this end, the TNTI was detected using a threshold of the kinetic energy defect \citep{Chauhan2014}. In the resulting figure \ref{figure8} histograms, the circle symbols indicate the detected peaks. The viscous-scaled segments reveal a greater number of UMZs on average, particularly at lower velocities corresponding to UMZs nearer to the wall. Using the average number of UMZs in panel \ref{figure8}(\textit{b}) rather than the smoother result in \ref{figure8}(\textit{a}) improves the eventual detection of these smaller UMZs. In this study, the histogram approach is preferred over kernel density estimation (KDE) to approximate the number of UMZs \citep{Fan2019}. The KDE method produces a smoothed estimate of the histogram distribution that can under-detect smaller UMZs in a manner similar to the large $\mathcal{L}_x$ result in figure \ref{figure8}(\textit{a}).

In the histogram method, the velocity associated with the UMZ interfaces can be taken as either the midpoint \citep{Adrian2000} or the minimum \citep{Heisel2018} between histogram peaks. Interfaces detected using the latter definition are shown in figure \ref{figure8}(\textit{c}) for the two $\mathcal{L}_x$ values. The TNTI is also included for reference. The isosurfaces for $\mathcal{L}_x^+=$ 2\,000 are determined independently within each segment such that the lines often end abruptly at the segment boundaries indicated by vertical dashed lines. The isosurfaces for $\mathcal{L}_x/\delta=$ 2 are continuous across the PIV field, but there are fewer detected UMZs, particularly in the lowest 20\% of the boundary layer which is the region of interest for the study.

The hybrid fuzzy clustering approach overcomes the limitations noted above by using the smaller $\mathcal{L}_x$ segments to estimate the number of UMZs, and the entire 2$\delta$-wide field to detect continuous UMZ interface isosurfaces. One important modification was made to the fuzzy clustering routine proposed by \citet{Fan2019}. An advantage of the histogram detection is that the shear profile separates the smaller UMZs closer to the wall from large UMZs in the wake. In other words, the smaller UMZs manifest a small distinct peak in lower-velocity histogram bins and are not suppressed by large peaks in higher-velocity bins. In contrast, the clustering algorithm considers neither the shear profile nor the ($x$, $z$) position of the velocity data. It was found that increasing the inputted number of clusters to the traditional algorithm often led to the division of large UMZs in the wake into two clusters rather than the detection of new cluster associated with a smaller UMZ. To ensure the detected clusters more closely reflect the histogram peaks, the PIV field was interpolated such that the resolution was proportional to the mean shear profile. The resolution of the interpolated grid was approximately ten times higher (i.e. lower $\Delta z$) near the wall and decreased with increasing $z$. The resolution profile was similar in magnitude and shape to non-uniform grids in numerical simulations. The interpolated grid effectively forced the algorithm to increase the weight of velocities with decreasing $z$, and significantly improved the detection of the smaller UMZs closer to the wall. The interpolation was only applied for the clustering algorithm; histogram peak detection and all other statistics were calculated on the original PIV vector field.

Finally, the clustering method was also used to detect the TNTI. Accordingly, the free stream region was included in the clustering algorithm, despite being excluded from the previous histograms. The inputted number of clusters was one greater than the average number of UMZs to account for presence of the free stream region. The  figure \ref{figure8}(\textit{c}) example includes the UMZ interfaces resulting from the clustering algorithm. There is general agreement between the TNTI detected using the kinetic energy and fuzzy clustering methods, noting that any differences do not affect the logarithmic region which is of most interest. Internal to the boundary layer, there is close overlap between the isosurfaces in the range $z/\delta=$ 0.4 to 0.6. The fuzzy clustering isosurfaces also align closely with many of the $\mathcal{L}_x^+=$ 2\,000 isosurface segments below $z/\delta=$ 0.3 that are undetected by the $\mathcal{L}_x/\delta=$ 2 histogram. The fuzzy clustering isosurfaces are more extensive, however, due to the continuity of the interfaces beyond the segment boundaries.

Many of the figures in the article were reproduced using the interfaces detected in $\mathcal{L}_x^+=$ 2\,000 segments. The results are qualitatively similar and the conclusions of the study do not differ between the two methods. The key advantage of the fuzzy clustering isosurfaces is the availability of results across wider distances to infer trends at the upper limit of the inertial subrange. While the hybrid detection was suitable for the present analysis, the method has not been tested for general use. For instance, it is not known whether the method would be appropriate for a significantly larger field of width $O(\gtrsim 10\delta )$. Future use of the same hybrid approach thus requires corroboration with alternate methods.

\section{Isolated pockets and the box-counting fractal dimension}
\label{appendixB}

For a given velocity level set, there are often numerous isolated isosurface pockets in addition to the longest isosurfaces that span the extent of the PIV field. With the present planar PIV measurements, it is not possible to determine whether each isolated pocket is connected to longer isosurfaces in the out-of-plane dimension, related to small-scale variability, or due to experimental noise. As discussed in \S \ref{section2_2}, the analysis in the main article excludes pockets whose area is smaller than $A_{min} = 10\lambda_T^2$. The dependence of the fractal dimension $D_j$ on $A_{min}$ is evaluated here. Two methods for incorporating the pockets into the average result are also discussed.

For the results in \S \ref{section3_1}, each isolated pocket is assumed to be independent of the long isosurfaces, and separate box counts are computed for each isosurface and pocket prior to ensemble averaging. In this case, an isolated pocket of length $O(\eta)$ would yield a constant value $N_b=1$ with slope $D_2=0$ for the range of $b/\eta$ in figure \ref{figure2}(\textit{a}). Numerous instances of these small isolated pockets would therefore bring $D_2$ closer to zero in the ensemble average of $N_b$. Removing these pockets ensures the overall length of the short isosurfaces does not bias the results. In this sense, the box counting results in figure \ref{figure2} represent only the longest isosurfaces where statistical self-similarity is apparent along the two-dimensional geometry of the surface.

\begin{figure}
\centerline{\includegraphics{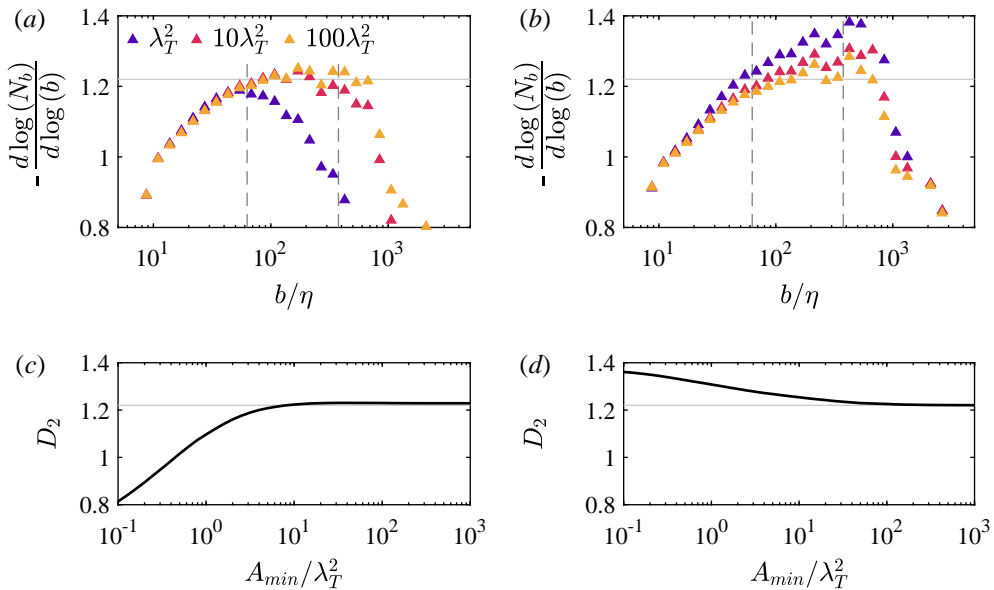}}
\caption{Sensitivity of the fractal dimension $D_2$ to the isolated pocket area threshold $A_{min}$ for $u_i=U(z=0.1\delta)$ and two averaging methods. In the left column (\textit{a,c}), separate box counts $N_b$ are computed for each pocket prior to averaging. In the right column (\textit{b,d}), a single box count is computed for all isosurfaces of $u_i$ in a given PIV field. Rows correspond to the local fractal dimension $D_2(b) = - d \log{(N_b)}/d \log{(b)}$ for three $A_{min}$ values (\textit{a,b}) where the vertical lines are the approximate limits of the inertial subrange and the average dimension $D_2$ within the inertial subrange as a function of $A_{min}$ (\textit{c,d}). The results in figure \ref{figure2} are based on separate box counts and $A_{min}=10\lambda_T^2$.}
\label{figure9}
\end{figure}

The fractal dimension $D_2(b) = - d \log{(N_b)}/d \log{(b)}$ resulting from the averaging method described above is shown for three values of $A_{min}$ in figure \ref{figure9}(\textit{a}). The primary effect of increasing $A_{min}$ is to remove the pockets with $N_b=1$ for large $b$, which increases the average value of $N_b(b)$ and enhances the region where $D_2$ is constant. The fractal dimension is approximately the same at the start of the inertial subrange. The dimension $D_2$ averaged within the inertial subrange is shown as a function of $A_{min}$ in figure \ref{figure9}(\textit{c}). The smaller $D_2$ values for smaller $A_{min}$ are due to decreasing $D_2(b)$ at the end of the inertial subrange apparent in figure \ref{figure9}(\textit{a}) for $A_{min}=\lambda_T^2$. The convergence of $D_2$ near $A_{min}=10\lambda_T^2$ supports the choice of this threshold for the main analysis. Larger thresholds exhibit the same trend of a constant value $D_2\approx$ 1.22 within the inertial subrange.

An alternate averaging approach is to assume the pockets are connected to the long isosurfaces along the out-of-plane $y$ direction. A single $N_b$ can then be estimated by grouping all isosurfaces and pockets of the same velocity $u_i$ within a given PIV field, and the ensemble average is computed across PIV realizations. Fractal dimension results from this averaging method are shown in figure \ref{figure9}(\textit{b,d}). The dimension $D_2(b)$ throughout the inertial subrange in this latter case exhibits a stronger dependence on $A_{min}$ than for the separated averaging case, and $D_2$ increases for decreasing $A_{min}$.

Both cases converge to the same $D_2$ value for large $A_{min}$ when all pockets are excluded and the difference in averaging methods is inconsequential. Modest scale dependence is apparent within the inertial subrange in both cases for small $A_{min}$, and the average $D_2$ appears increasingly constant for larger $A_{min}$. The conclusion in the main article regarding a constant fractal dimension throughout the inertial subrange is therefore independent of the averaging method, so long as a sufficiently large filter $A_{min}$ is employed to focus the analysis on the largest continuous isosurface geometries.

\section{Isosurface anisotropy and the box-counting fractal dimension}
\label{appendixC}

\begin{figure}
\centerline{\includegraphics{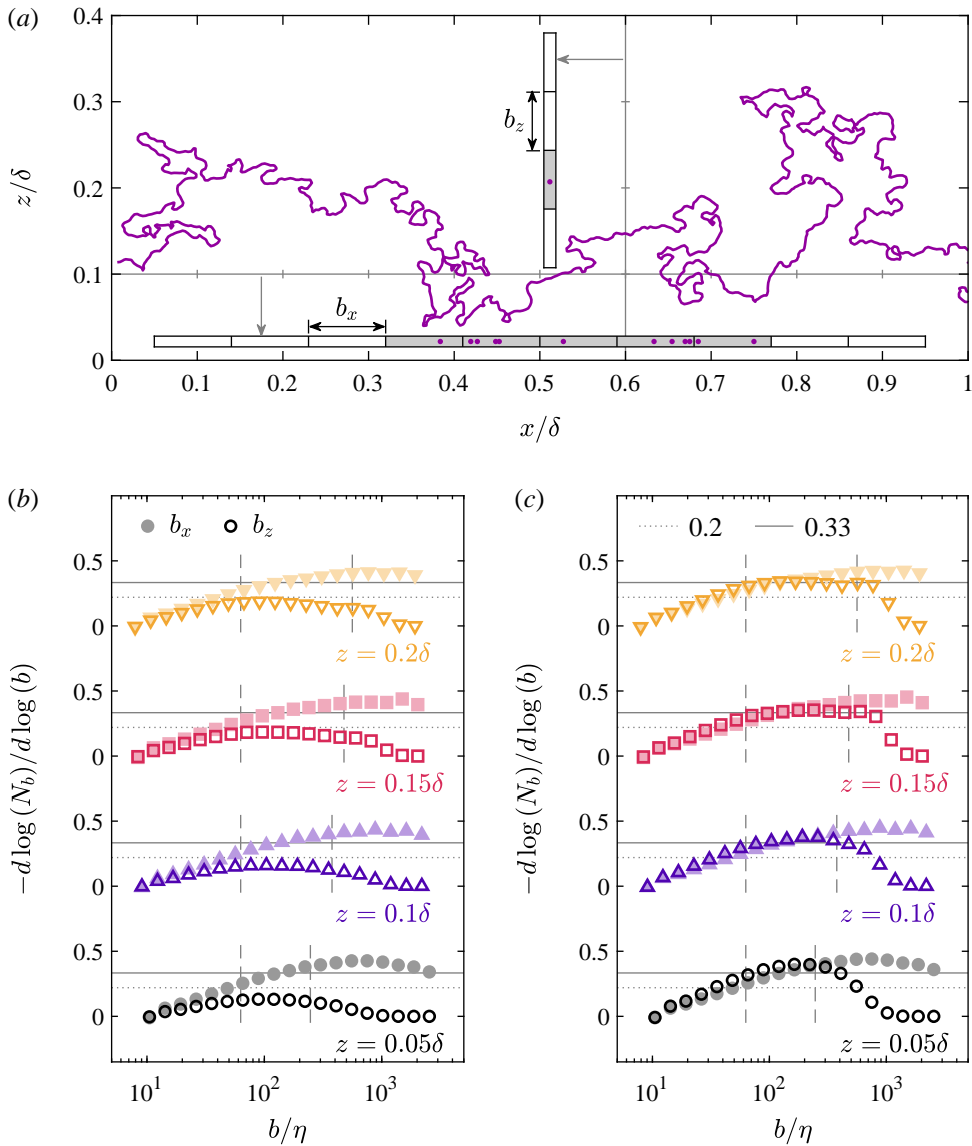}}
\caption{Estimates of the lower-order fractal dimension $D_1$ based on box counts along streamwise and wall-normal transects of the velocity isosurfaces. (\textit{a}) An example isosurface and one-dimensional transects (gray lines). The insets show the number of transect segments $N_b$ (shaded) of length $b=0.1\delta$ containing intersections with the isosurface (dots). (\textit{b}) Local fractal dimension $D_1(b) = - d \log{(N_b)}/d \log{(b)}$ resulting from box counts along $x$ and $z$ for varying $b$. (\textit{c}) Local fractal dimension when transects with a single intersection yielding $N_b(b)=$ 1, e.g. the example wall-normal transect in (\textit{a}), are excluded from the averaged result. The vertical dashed lines are the approximate limits of the inertial subrange.}
\label{figure10}
\end{figure}

An additional aspect of the box-counting fractal dimension explored here is the difference between the value $D_2 \approx$ 1.2 observed in \S \ref{section3_1} and the approximate value 1.33 previously reported for the TNTI \citep{Sreenivasan1986,deSilva2013,Chauhan2014,Borrell2016}. Directional trends in the box-counting result can be inferred by conducting separate one-dimensional counts along the streamwise and wall-normal directions. In the streamwise case, a transect of the isosurface is taken across a given measured $z$ position, and the $x$ positions where the isosurface intersects the transect are compiled. The transects are discretized into segments of length $b_x$ and the number of segments $N_b$ containing at least one crossing is counted. The process is repeated across $z$ positions. The same principle is used to compute $N_b$ for wall-normal transects of the isosurfaces. An example of the one-dimensional count is shown in figure \ref{figure10}(\textit{a}), and the resulting fractal dimensions $D_1(b)$ along $x$ and $z$ are given in figure \ref{figure10}(\textit{b}). If a self-similar geometry is isotropic and independent transects are taken to estimate a lower-order fractal dimension \citep[see, e.g., figure 2 of][]{Sreenivasan1986}, the fractal dimensions are related as $D_2 =D_1+1$ \citep{Mandelbrot1982}.

The fractal dimension for streamwise transects (filled data markers) appears scale dependent even within the inertial subrange. This dependence has been observed for previous one-dimensional estimates of fractal behavior \citep{Miller1991,Praskovsky1993}, and is a stochastic consequence of a finite-sized power law (i.e. finite Reynolds number) \citep{Catrakis2000,Heisel2022}. However, the approximate dimension $D_1 \approx$ 0.33 near the center of the inertial subrange matches the value observed in previous studies for the TNTI.

Interestingly, the dimension for wall-normal transects (open data markers) is less variable within the inertial subrange but has a consistently smaller value $D_1 \approx$ 0.15 throughout the logarithmic region. The discrepancy in $D_1$ values can be explained by the occurrence of transects with a single isosurface crossing, e.g. the example wall-normal transect in figure \ref{figure10}(\textit{a}). Similar to the $O(\eta)$ pockets discussed in appendix \ref{appendixB}, a single crossing represents a trivial case that yields a constant value $N_b(b)=1$ with slope $D_1=0$ regardless of box size, and thus reduces $D_1$ in the ensemble average. For the transects contributing to figure \ref{figure10}(\textit{b}), 65--80\% of wall-normal transects have a single crossing compared to only 5--7\% of streamwise transects.

Figure \ref{figure10}(\textit{c}) shows the fractal dimension when these trivial transects are excluded from the average result. The dimension for streamwise transects is approximately unchanged, but the wall-normal case is now in close agreement with $D_1 \approx$ 0.33 such that both directions exhibit the same fractal behavior. The frequent occurrence of trivial wall-normal transects is likely due to large-scale anisotropy in the shape of the isosurfaces, where the largest features within the isosurface are longer in $x$ than $z$. This anisotropy is consistent with the observed aspect ratio of large-scale coherent structures \citep[e.g.,][]{delAlamo2006,Lozano2012,Baars2017}, and is reflected by the difference in $D_1(b)$ in figure \ref{figure10}(\textit{c}) for the largest box sizes exceeding the inertial subrange.

We speculate that the large-scale anisotropy of the isosurfaces similarly contributes to the two-dimensional result $D_2 \approx$ 1.2. The anisotropy leads to more sparse regions along $z$ than along $x$. The average results of the box counting methodology include these sparse regions, which may reduce the average dimension $D_2$ in the same way that $D_1$ is reduced in figure \ref{figure10}(\textit{b}). Further investigation is therefore warranted before conclusions are made regarding the observed value $D_2\approx$ 1.2 for boundary layer turbulence. Importantly, the trivial regions with $N_b(b)=1$ affect all box sizes in the same manner and do not change the range of sizes $b$ exhibiting a power law in the ensemble average. While the anisotropy is expected to reduce the observed value for $D_2$, it does not influence the primary conclusion in \S \ref{section3_1} regarding a constant fractal dimension within the inertial subrange.

\bibliographystyle{jfm}
\bibliography{references}

\end{document}